\documentclass[times,twocolumn,final]{elsarticle}
\usepackage{medima}
\usepackage{framed,multirow}

\usepackage{amssymb}
\usepackage{latexsym}

\usepackage{url}
\usepackage{xcolor}

\usepackage{hyperref}

\definecolor{newcolor}{rgb}{.8,.349,.1}

\journal{arXiv.org}

\begin{document}

\verso{Georg Hille \textit{et~al.}}

\begin{frontmatter}

\title{Joint liver and hepatic lesion segmentation in MRI using a hybrid CNN with transformer layers}%

\author[1]{Georg \snm{Hille}\corref{cor1}}
\cortext[cor1]{Corresponding author: 
  Tel.: +49391-67-57272}
  \ead{georg.hille@ovgu.de}
\author[1]{Shubham \snm{Agrawal}}
\author[1]{Pavan \snm{Tummala}}
\author[2]{Christian \snm{Wybranski}}
\author[2]{Maciej \snm{Pech}}
\author[2]{Alexey \snm{Surov}}
\author[1]{Sylvia \snm{Saalfeld}}
%% Third author's email

\address[1]{Department of Simulation and Graphics, Otto-von-Guericke University, Magdeburg, Germany}
\address[2]{Department of Radiology, University Hospital of Magdeburg, Magdeburg, Germany.}

\begin{abstract}
%%%
\textbf{Backgound and Objective:} Deep learning-based segmentation of the liver and hepatic lesions therein steadily gains relevance in clinical practice due to the increasing incidence of liver cancer each year. Whereas various network variants with overall promising results in the field of medical image segmentation have been successfully developed over the last years, almost all of them struggle with the challenge of accurately segmenting hepatic lesions in magnetic resonance imaging (MRI). This led to the idea of combining elements of convolutional and transformer-based architectures to overcome the existing limitations. 
\textbf{Methods:} This work presents a hybrid network called SWTR-Unet, consisting of a pretrained ResNet, transformer blocks as well as a common Unet-style decoder path. This network was primarily applied to single-modality non-contrast-enhanced liver MRI and additionally to the publicly available computed tomography (CT) data of the liver tumor segmentation (LiTS) challenge to verify the applicability on other modalities. For a broader evaluation, multiple state-of-the-art networks were implemented and applied, ensuring a direct comparability. Furthermore, correlation analysis and an ablation study were carried out, to investigate various influencing factors on the segmentation accuracy of the presented method. 
\textbf{Results:} With Dice similarity scores of averaged $98~\pm~2~\%$ for liver and $81~\pm~28~\%$ lesion segmentation on the MRI dataset and $97~\pm~2~\%$ and $79~\pm~25~\%$, respectively on the CT dataset, the proposed SWTR-Unet proved to be a precise approach for liver and hepatic lesion segmentation with state-of-the-art results for MRI and competing accuracy in CT imaging. 
\textbf{Conclusion:} The achieved segmentation accuracy was found to be on par with manually performed expert segmentations as indicated by inter-observer variabilities for liver lesion segmentation. 
In conclusion, the presented method could save valuable time and resources in clinical practice.
%%%%
\end{abstract}


\end{frontmatter}

%\linenumbers

\section{Introduction}
\label{sec:introduction}

Following cardiovascular diseases, cancer constitutes the second major cause of death, accounting for about 8.8 million deaths worldwide in 2015, with the liver being one of the most common sites for the development of primary and metastatic lesions \citep{b0}. 
%With regards to diagnosis and disease progression, abnormal shapes and textures of the liver, as well as present lesions visible in medical imaging represent relevant biomarkers \cite{b1}. 
Besides metastatic lesions, which often originate from primary breast, colon and pancreas cancer, the liver is also site of primary tumor development \citep{b2}. The Hepatocellular carcinoma (HCC) is among the most frequent tumor variants and causes the third-most cancer-related deaths worldwide with a growing incidence over the last decades \citep{b2}. In terms of diagnosis and therapy planning, medical imaging of the liver plays a vital role, either during the routinely performed tumor staging of primary lesions outside of the liver or if the clinical anamnesis points towards primary hepatic cancer diseases. For such imaging purposes, magnetic resonance imaging (MRI) and computed tomography (CT) are obligatory, each with its own benefits and disadvantages. \
%While CT offers high spatial resolution and minimal susceptibility for motion artifacts due to fast acquisition times, the major drawbacks lie in the radiation exposure and the inferior soft tissue contrast compared to MRI. The latter enables freely adjustable cross-section angles and varying tissue contrasts due to multiple acquisition sequences, which pronounce different physiological tissue compositions. These advantages come with the cost of time-consuming acquisition procedures, which commonly limit the overall spatial resolution and are prone to multiple image artifacts, e.g., motion or bias artifacts. \

The precise identification and segmentation of hepatic lesions in medical imaging could support radiologists in tumor staging and therapy decision-making. In current clinical routine it is common that such segmentation procedures are performed manually, which represents the gold standard or semi-automatically with algorithmical support. Either way, both strategies and particularly manually contouring, are time-consuming, cumbersome, operator-dependent and subjective. Even more automated approaches, which have been developed in the past, require manual initialization and their segmentation accuracy often heavily rely on the precision of said initial user interactions. Despite that, it has been shown that precisely segmenting hepatic lesions is a highly challenging task on its own due to the vast variability of shape, texture, size, location and number of liver lesions per patient case \citep{b25}. The tissue image contrast between liver and lesion highly varies depending on the used acquisition protocols or the application of contrast-agents. Thus, it is hardly possible to define a model-based segmentation approach based on crafted 
features by a priori knowledge. \
Such approaches include graph cuts, level sets or clustering techniques, that have been applied almost exclusively to liver CT data \citep{b3,b4,b5}. With respect to the imaging modality, the same holds true for machine or deep learning-based methods, which were almost exclusively applied to liver CT images. Nevertheless, learning-based approaches seem to be the most promising strategy to deal with the sheer appearance variability of liver lesions \citep{b20}. In recent years various works focused on CT-based liver and liver lesion segmentation using fully convolutional neural networks (FCNN) \citep{b13,b35,b36}, especially since the liver tumor segmentation (LiTS) challenge publicly provided patient CT scans, which are otherwise hard to compromise in reasonable numbers. 
In terms of MRI-based hepatic lesion segmentation, there currently are, to the best of the authors' knowledge, only a few works of direct relevance, such as from \cite{b6}, \cite{b48}, \cite{b49} and \cite{b50}. \cite{b6} presented a fully automatic approach consisting of a cascaded convolutional neural network, that first segments the liver outline and subsequently the hepatic lesions within the resulting liver mask images. The two-step strategy was applied to both, liver CT and a few MR images \citep{b6}. \cite{b48} applied an U-net with post-processing to minimize the false positive rate to multiphasic contrast-enhanced MR images. \cite{b49} presented a united adversial learning framework for hepatic lesion segmentation and detection of multimodalty non-contrast-enhanced MRI. Their network consists of a multimodal feature extraction encoder of three parallel convolutional channels, each incorporating a specific MRI sequence, followed by a feature fusion and selection step, whereas the paths of both sub-tasks, i.e., segmentation and detection share coordinates. The results of both paths are subsequently fed into a multiphase radiomics guided discriminator to improve performance by adversial learning. \cite{b50} applied an anisotropic 3D Unet with multi-model training to contrast-enhanced liver MRI. In order to avoid unfavourable random weight initiations of the network, the multi-model approach started to train $2^4$ models in parallel, dismissing half of it according to the validation Jaccard scores after a fix number of iterations. 
Most of the above-mentioned works utilized convolutional neural networks (CNN) in some form, which proved to be superior to many methodological alternatives with respect to their representational performance. However, their ability to model long-range relations showed limitation due to the locality of convolutional operations \citep{b16}. This may be crucial when it comes to structures with high appearance variations like hepatic lesions. In that regard, techniques like attention mechanisms were introduced to support CNN-based architectures \citep{b37,b16} or widely replace convolutional blocks in form of transformer elements \citep{b26}. \

% and achieved averaged liver Dice similarity scores of $94\%$ (CT) and $87\%$ (MRI), as well as lesion Dice similarity scores on average of $61\%$ (CT) and $69.7\%$ (MRI). The dataset they used was comprised of 100 CT patient scans and thereof 20 validation cases as well as 31 MRI scans without any details about the validation set size. 
%In contrast to MRI-based hepatic lesion segmentation, various works dealing with CT imaging have been presented in recent years. Both, Unet-style architectures processing 2D image input like \cite{b7,b8,b9} and 3D-based approaches like \cite{b10,b11} and even hybrid variants, which combine 2D and 3D components \cite{b12} have been applied to this particular segmentation task. Isensee et al. \cite{b35} proposed an U-net-based approach applicable to various segmentation tasks without the need of task-specific adaption or manually fine-tuning most of the hyperparameters. 
%Fan et al. \cite{b13} introduced attention mechanisms, originated from the computer vision domain and included such block-wise elements within a common CNN-architecture. 
%Zhang et al. \cite{b36} proposed a network incorporating cross-phase information of multi-phase CT imaging, as well as local uncertainty to deal with lesion boundaries that are hard to trace.
%Although, all of these works achieved liver segmentation accuracies of $>90~\%$ and most of them even $>95~\%$ Dice, they commonly lack sufficient segmentation quality regarding the hepatic lesions, which therefore represents and remains the primary challenge. \ 

The aim of this work was to develop a fully automatic deep learning-based joint liver and hepatic lesion segmentation approach, which incorporates attention mechanisms to both, achieve expert-like segmentation accuracy and with applicability to clinical MRI and CT data. Furthermore, to ensure a more direct comparability of the proposed method with relevant works from the state of the art, multiple network variants were additionally implemented and tested on the same imaging data. 

\subsection{Related Work}

\noindent \textbf{CNN-based methods.} In recent years, machine or deep learning-based approaches achieved vastly superior results in terms of medical image segmentation tasks, representing almost exclusively the state of the art in this field. Especially the introduction of the Unet \citep{b27} led to a dominance of U-shaped variants like ResUnet \citep{b37}, DenseUnet \citep{b32} or nnUnet \citep{b35}. However, there were various works using an Unet-styled network as a basic architecture, whereas different adaptions were made to increase the performances of CNNs. For instance, sequential models were embedded as CNN bottlenecks \citep{b46,b47} to tackle imbalanced data sets or for joint detection and segmentation purposes.  

\noindent \textbf{CNNs with self-attention or transformer-based elements.}
In order to sufficiently capture global information and overcome the limitation of CNNs due to their intrinsic locality of convolutional operations, various self-attention mechanisms were introduced \citep{b39,b40}. Whereas these approaches widely follow the principal architecture of U-shaped CNNs, some works proposed to combine convolutional- with transformer-based elements more recently \citep{b16,b41}.
Transformers were first introduced in the field of natural language processing and quickly established itself as a state-of-the-art approach \citep{b18}. Adaptions like sparse transformers \citep{b42} or local self-attention \citep{b43} enabled the applicability to computer vision tasks. So called vision transformers showed excellent results in image recognition tasks, but come with the cost of pretraining on large data sets on its own, before fine-tuning on domain-specific data, when compared to CNN-based approaches. However, multiple training strategies were proposed to exploit the benefits of ImageNet in that regard \citep{b44}. Such a pretrained transformer could be embedded within the encoder path of an U-net-styled architecture for medical image segmentation \citep{b16, b45}.
Furthermore, SWIN transformers were introduced, which represent very efficient hierarchical vision transformers based on shifted window mechanisms \citep{b18}. Since SWIN transformers showed very promising results on multiple vision tasks including semantic segmentations, they were utilized as the basic unit in a purely transformer-based U-shaped network architecture called SWIN-Unet \citep{b26}. In contrast to the above-mentioned approaches, the presented work combines the most promising elements of CNNs and transformer-based self-attention mechanisms, resulting in a hybrid CNN-transformer-based architecture using SWIN transformer blocks.

\begin{figure}[t]
  \centering
\includegraphics[width=\columnwidth]{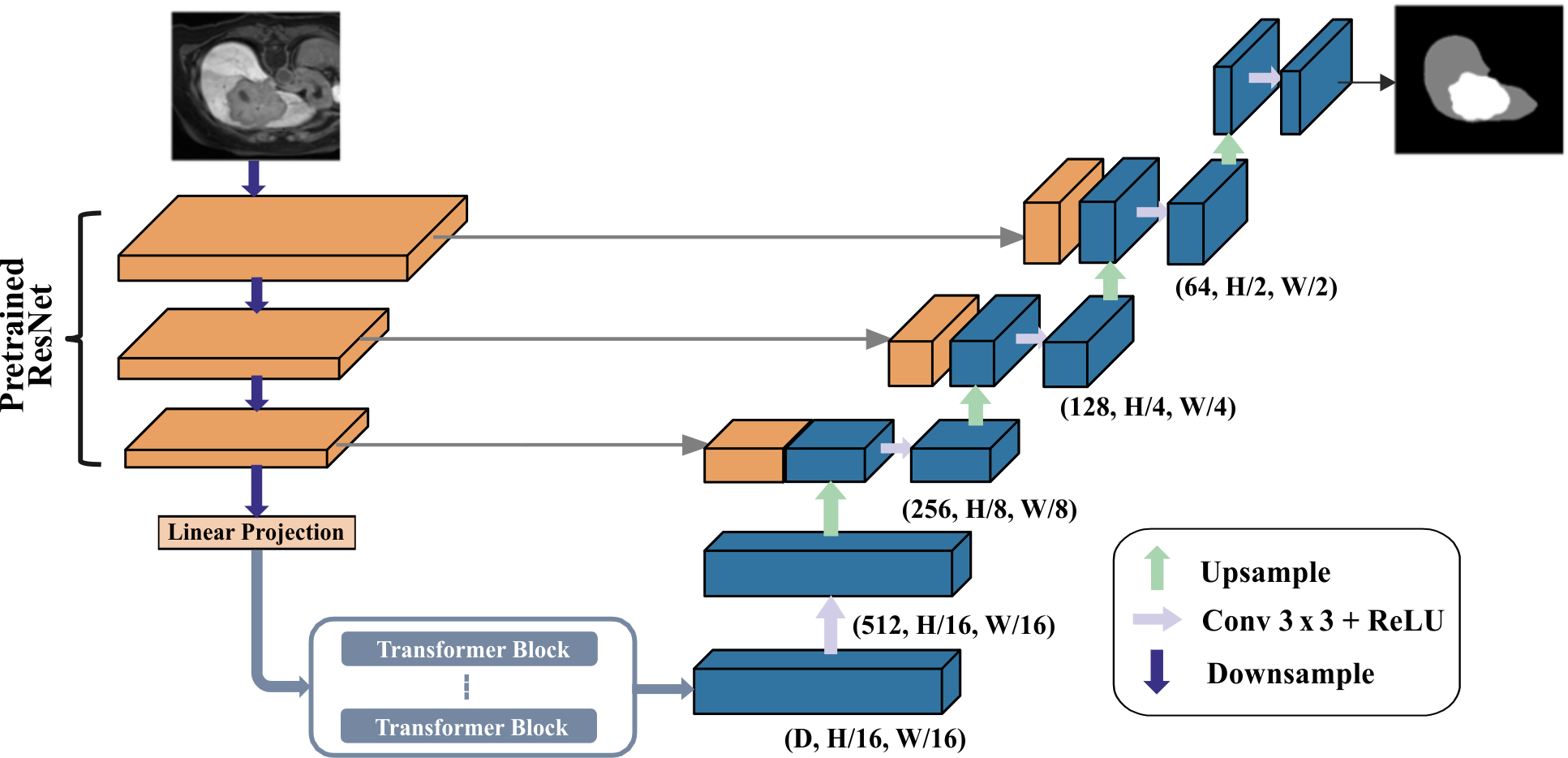}
\caption{The basic architecture of the proposed hybrid SWTR-Unet, that combines a ResNet-styled encoder path, transformer blocks at the Unet-bottleneck and a common convolutional decoder path.}
\label{figNet}
\end{figure}

\

\section{Methods}

\begin{figure}[t]
\includegraphics[width=\columnwidth]{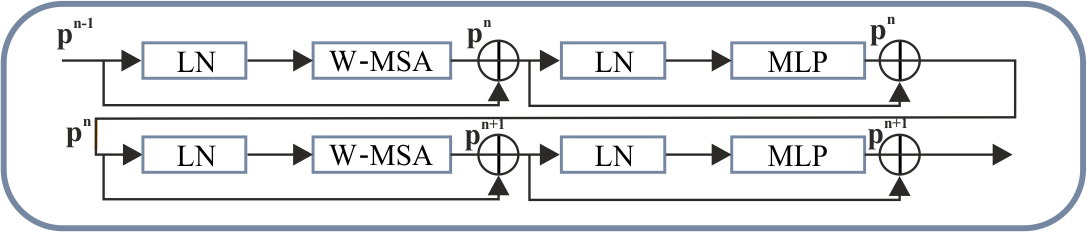}
\caption{Structure of a SWIN-transformer block proposed by \cite{b26} consisting of a sequence of LayerNorm (LN) layers, window-based multi-head self-attention (W-MSA) or shifted window-based multi-head self-attention (SW-MSA) blocks and multilayer perceptrons (MLP).}
\label{figtrans}
\end{figure}

Figure \ref{figNet} and \ref{figtrans} show the architecture of the proposed SWTR-Unet (SWIN-transformer-Unet) network for joint liver and hepatic lesion segmentation in MRI and CT. Herefore, Res-blocks, transformer-based multi-head self-attention (MSA) blocks as well as shifting window (SWIN) transformer blocks were combined into a hybrid structure. The components of the SWTR-Unet will be described in more detail in the following.

\subsection{SWTR-Unet}

Regarding the hybrid encoder design of the proposed network a ResNet \citep{b14} was implemented as a convolutional backbone, which was pretrained on ImageNet \citep{b15} and combined with in total twelve transformer blocks at the bottleneck of the U-shaped architecture. Beside the function as a robust feature extractor, the Res-blocks within the pretrained ResNet also address the vanishing gradient problem and allow a more dynamic and efficient learning \citep{b14}. Fully convolutional-based networks like the widely known and applied Unet commonly suffer from the limitation of capturing long-range dependencies due to the intrinsic locality of convolutional operations. For extracting the global contextual information, the transformer elements encode the input images as a sequence of image patches. Then, the decoder performs the upsampling task needed for retrieving the precise localization. The transformer architecture replaces convolutional operators and relies on a multi-head self-attention mechanism instead. In contrast to works like \citep{b16,b26}, the twelve transformer layers in the SWTR-Unet consist of multiple consecutive SWIN transformer sub-blocks, forming a SWIN transformer-based bottleneck within a CNN-based architecture. Each of those sub-blocks comprises layer normalization (LN), multi-headed self-attention (MSA) modules, residual connections, and a 2-layer multilayer perceptron (MLP) with a Gaussian Error Linear Unit (GELU) activation for adding non-linearity to the network. Regarding the self-attention modules, window-based multi-headed self-attention (W-MSA) and shifted window-based multi-headed self-attention (SW-MSA) are successively applied.

The input for the SWIN transformer blocks as the Unet bottleneck is generated by the preceding linear projection layer, which reshapes the 2D image in a sequence of 2D image tokens or patches, respectively of size 16 $\times$ 16 pixels \citep{b17}. The transformer architecture conducts global self-attention by computing the relationship of each patch with all other patches of the fed sequence, resulting in a computation of quadratic complexity \citep{b18}. This would make global self-attention mechanisms unsuitable for larger images. In order to overcome this limitation, shifted window-based local self-attention as proposed by Vaswani et al. \citep{b18}, could reduce the computational costs due to linear complexity and furthermore, introduce cross-window connections via shifting.

The final segmentation masks of liver and lesions are obtained by the decoder-path of the STWR-Unet, which consists of in total four up-sampling layers. Starting at the bottom of the U-shaped architecture, the sequence of hidden features as output of the transformer blocks, is reshaped and then successively led through the decoder layers, where each consists of a two-times upsampling layer, a 3 $\times$ 3 convolutional layer and a Rectified linear unit (ReLU) layer. Therefore, the proposed SWTR-Unet is capable of aggregating both, local and global features as present in different resolution levels by preserving and transferring them to the decoder side via skip connections. Finally, the feature map is passed through the segmentation head to generate the segmentation output.

\begin{table*}[t!]
  \centering
  \caption{Hyperparameter settings for each of the tested networks, both the re-implemented state-of-the-art architectures and the proposed SWTR-Unet variant. The center line separates pure convolutional networks from those with transformer elements. ReLu - Rectified Linear Unit, GeLu - Gaussian Error Linear Unit, SGD - Stochastic Gradient Descent, RMSProp - Root Mean Square Propagation, BCE - Binary Cross-entropy, CE - Cross-entropy, Params - Number of trainable parameters.}
 \setlength{\tabcolsep}{7pt}
\begin{tabular}{ccccc ccccccc}

\multicolumn{6}{c}{ }\\
\hline
Network & Activation Function & Optimizer & Loss Function & Epochs & Learning rate & Params\\
\hline
%& \\
Unet \citep{b27}       & ReLU       & Adam    & BCE       & 60 & 1e-3 & 8M\\
DeepLabV3  \citep{b29} & ReLU       & Adam    & BCE       & 60 & 1e-4 & 63M\\
Attn. Unet \citep{b28} & Leaky ReLU & SGD     & BCE       & 60 & 1e-3 & 6M\\
PSPNet \citep{b30}     & Leaky ReLU & RMSProp & Dice      & 70 & 1e-4 & 68M\\
DAF3D  \citep{b21}     & Leaky ReLU & Adam    & Dice + CE & 70 & 1e-4 & 29M\\
DenseUnet \citep{b32}  & Leaky ReLU & Adam    & Dice + CE & 70 & 1e-3 & 36M\\
nnUnet \citep{b35}     & Leaky ReLU & SGD     & Dice + CE & 1000 & adaptive & 30M\\
\hline      
UnetR \citep{b31}      & ReLU       & SGD     & Dice      & 60 & 1e-4 & 93M\\
SwinUnet \citep{b26}   & GeLU       & Adam    & Dice      & 80 & 1e-4 & 100M\\
TransUnet \citep{b16}  & GeLU       & Adam    & Dice + CE & 80 & 1e-4 & 97M \\
\textbf{SWTR-Unet}  & \textbf{GeLU} & \textbf{SGD} & \textbf{Dice + CE} & \textbf{70} & \textbf{1e-4} & \textbf{110M}\\
\hline
\end{tabular}%
\label{tab:para}
\end{table*}

\subsection{Imaging Data}

Since both, MR and CT imaging are of crucial importance in clinical routine of hepatic lesion diagnosis and treatment decision-making, the same pipeline of pre-processing, training and evaluation was applied to both imaging modalities. 

The single-modality, monophasic non-contrast-enhanced MR imaging data was retrospectively compiled and originally used for intervention planning purposes before brachytherapy at the Department of Radiology and Nuclear Medicine of the University Hospital in Magdeburg. The images were acquired using a Philips Intera 1.5~T scanner and an eTHRIVE sequence with repetition (TR) and echo times (TE) of 4.0 - 4.1~ms and 2.0~ms, respectively. In-plane image resolution of the axial scans was 0.98 mm, the spacing between slices was 3~mm, resulting in acquisition matrices of size $224 \times 224 \times 64$. A total of 48 patient cases with overall 157 hepatic lesions were compiled. The ground truth segmentations of the liver outline and the hepatic lesions therein were performed by an experienced radiologist.

With respect to CT imaging, the LiTS challenge \citep{b20} data was used, which comprises scans from several clinical sites, using different acquisition protocols and CT scanners, and thus, the image quality and resolution vary noticeably. In-plane image resolution of the axial scans varies from 0.56~mm to 1.0~mm and the slice thickness varies from 0.45~mm to 6.0~mm. The number of slices per volume ranges from 42 to 128. The hepatic lesions that are present in each patient scan vary in size between 38~mm$^3$ and 349~cm$^3$ \citep{b20}.

\subsection{Pre-processing and Augmentation}

The following description of the data pre-processing and augmentation holds true for both the MRI and CT data, if not stated otherwise. First, adaptive histogram equalization was performed to enhance the contrast volume-wise followed by resampling to a fixed matrix size. With regard to the MRI data, z-score normalization was performed, followed by an N4 bias correction. For CT data, the Hounsfield units were limited to the range [-100; 400] to exclude irrelevant outlier pixel intensities and subsequently normalized using the 5th and 95th percentile of the foreground intensities. 
Data augmentation consisting of various intensity and geometrical approaches were employed to strengthen the network's robustness and generalization ability, since the number of available patient cases was relatively small, especially compared to the number of trainable parameters of each network. The image data was augmented using the application of Gaussian noise, gamma and affine transformations, including flipping (overall probability of $60\%$, for each direction $50\%$), rotation ($\pm20^{\circ}$) and translation ($\pm32$ voxels for the x and y-axis and $\pm16$ voxels for the z-axis). Each of the 48 volumes of the MRI dataset was augmented 20-times leading to 960 samples of size $224 \times 224 \times 64$ voxels. Similarly, the 131 CT volumes were augmented 20-times and yielded in total 2,620 samples. Both, data sets were split into single slices for 2D image input, resulting in a total of 61,440 MR and 189,600 CT images.

\subsection{Implementation details}

The implementation of the proposed SWTR-Unet as well as relevant state-of-the-art methods were realized with Python 3.6 and Pytorch 1.7.0. All of the models were trained on a NVIDIA GeForce RTX2080 Ti GPU with 12GB of memory. Due to the limitation of GPU memory, the used batch size was 32 for 2D input and 8 in the case of the 3D input of the DAF3D network \citep{b21}. Table \ref{tab:para} describes the hyperparameter used after fine-tuning for each of the experimented methods.
In order to evaluate the capability of the methods listed in Table \ref{tab:para} to segment the liver and lesions therein, seven-fold-cross validation was performed. Therefore, the MR dataset was separated into seven subsets each containing the augmented data of 41 to 42 image volumes for training purposes as well as six to seven original volumes within a corresponding validation set. Similarly, the CT data set is separated into seven-fold, each containing the augmented data of 114 image volumes for training purposes as well as 17 original volumes within a corresponding validation set. In doing so, it is guaranteed that the networks produce segmentation masks on unseen images with respect to the training process per fold. The 2D networks were applied to all slices per validation volume and predicted 2D slices were subsequently combined patient-wise to calculate the quality metrics regarding the segmentation accuracy per patient image volume. The results stated in the following refer to the average over all seven folds. In order to assess and compare the segmentation performance of all networks, no further post-processing was implemented.

\begin{table*}[t!]
  \centering
  \caption{Experimental results produced on the MRI data for each of the tested networks including the proposed SWTR-Unet. Stated are the Dice similarity coefficients (DSC) and Hausdorff distances (HD) of the liver and liver lesion segmentations averaged over all seven folds. The center line separates pure convolutional networks from those with transformer elements.}
 \setlength{\tabcolsep}{7pt}
\begin{tabular}{ccccc ccccccc}

\multicolumn{6}{c}{ }\\
\hline
& \multicolumn{2}{c}{DSC} & & \multicolumn{2}{c}{HD}\\
& $DSC_{liver}$ & $DSC_{lesion}$ & & $HD_{liver}$ & $HD_{lesion}$ \\
\hline
%& \\
Unet \citep{b27}       &       $0.93 \pm 0.06$ & $0.36 \pm 0.38$ & & $6.52 \pm 9.03$ & $5.49 \pm 14.07$ \\
DeepLabV3  \citep{b29} & $0.87 \pm 0.10$ & $0.36 \pm 0.33$ & & $8.35 \pm 10.50$ & $9.97 \pm 11.66$ \\
Attn. Unet \citep{b28} & $0.91 \pm 0.09$ & $0.64 \pm 0.29$ & & $9.74 \pm 9.82$ & $10.07 \pm 7.32$ \\
PSPNet  \citep{b30}    & $0.95 \pm 0.01$ & $0.67 \pm 0.32$ & & $3.04 \pm 5.20$ & $17.26 \pm 25.82$ \\
DAF3D  \citep{b21}     & $0.87 \pm 0.14$ & $0.73 \pm 0.23$ & & $18.66 \pm 25.43$ & $9.00 \pm 22.03$ \\
DenseUnet \citep{b32}  & $0.96 \pm 0.01$ & $0.74 \pm 0.23$ & & $1.30 \pm 0.98$ & $14.11 \pm 26.52$ \\
nnUnet \citep{b35}    & $0.97 \pm 0.02$ & $0.80 \pm 0.21$ & & $4.30 \pm 1.11$ & $8.32 \pm 13.10$ \\
\hline      
UnetR    \citep{b31}   & $0.90 \pm 0.07$ & $0.49 \pm 0.26$ & & $66.49 \pm 49.41$ & $34.26 \pm 41.81$ \\
SwinUnet  \citep{b26}  & $0.91 \pm 0.14$ & $0.76 \pm 0.30$ & & $6.32 \pm 4.72$ & $4.71\pm 6.77$ \\
TransUnet \citep{b16}  & $0.97 \pm 0.03$ & $0.76 \pm 0.27$ & & $1.05 \pm 0.23$ & $11.56 \pm 25.54$ \\
\textbf{SWTR-Unet}  & $\mathbf{0.98 \pm 0.02}$ & $\mathbf{0.81 \pm 0.28}$ & & $\mathbf{1.02 \pm 0.18}$ & $\mathbf{7.03 \pm 17.37}$ \\
\hline
\end{tabular}%
\label{tab:mri}
\end{table*}
\

\subsection{Quality metrics}

In order to assess and compare the capability of liver and hepatic lesion segmentation of the proposed SWTR-Unet with the state-of-the-art networks, Dice similarity coefficients (DSC) and Hausdorff distances (HD) per liver and the lesions within were used. The former is defined as

\begin{equation}DSC=\frac{2|X \cap Y|}{|X| + |Y|}, \label{dice}\end{equation} 

where $|X|$ and $|Y|$ represent the reference and the resulting network's segmentation. The Hausdorff distance, also known as the maximum surface distance, is defined as:

\begin{equation}HD(X,Y)= max\{\hat{H}(X,Y),\hat{H}(Y,X)\}~~~with \label{HD}\end{equation} 
\begin{equation}\hat{H}(X,Y)= max\{min\{|x,y|\}\} \label{dHD}\end{equation} 

being the maximum of both directed Hausdorff distances $\hat{H}$ between the two surface point sets $X$ of the reference and $Y$ of the produced segmentation in each direction. $\hat{H}$ was the maximum distance between any point x $\in$ X and its nearest neighbour y $\in$ Y and therefore, represents the worst contour misalignment.

\section{Results}

The evaluation of the proposed SWTR-Unet consists of three different aspects, which will be discussed successively. First of all, the SWTR-Unet as well as multiple state-of-the-art approaches that were additionally re-implemented, were applied to the MRI data. Hence, the segmentation capability of all network variants regarding the liver and its hepatic lesions therein could be compared directly. The second part focuses on the application of the proposed approach to the CT imaging data of the LiTS challenge and therefore, the segmentation accuracy can be assessed on both of the most relevant liver cancer imaging modalities. The final part was conducted to get more insights on potential influencing factors on the segmentation accuracy of the proposed SWTR-Unet, e.g., the number of skip connections and transformer layers or how lesion size or shape influence the segmentation results.
Additionally, for assessing the quality of the segmentation results, a comparison with the inter-observer variability (IOV) of hepatic lesion segmentations manually produced by experts is instructive. Related literature stated IOV values as Dice scores of $78 \pm 12\%$ for MR imaging \citep{b50} and $64-82\%$ produced on CT images \citep{b24}. 

\begin{figure}[!t]
\centerline{\includegraphics[width=\columnwidth]{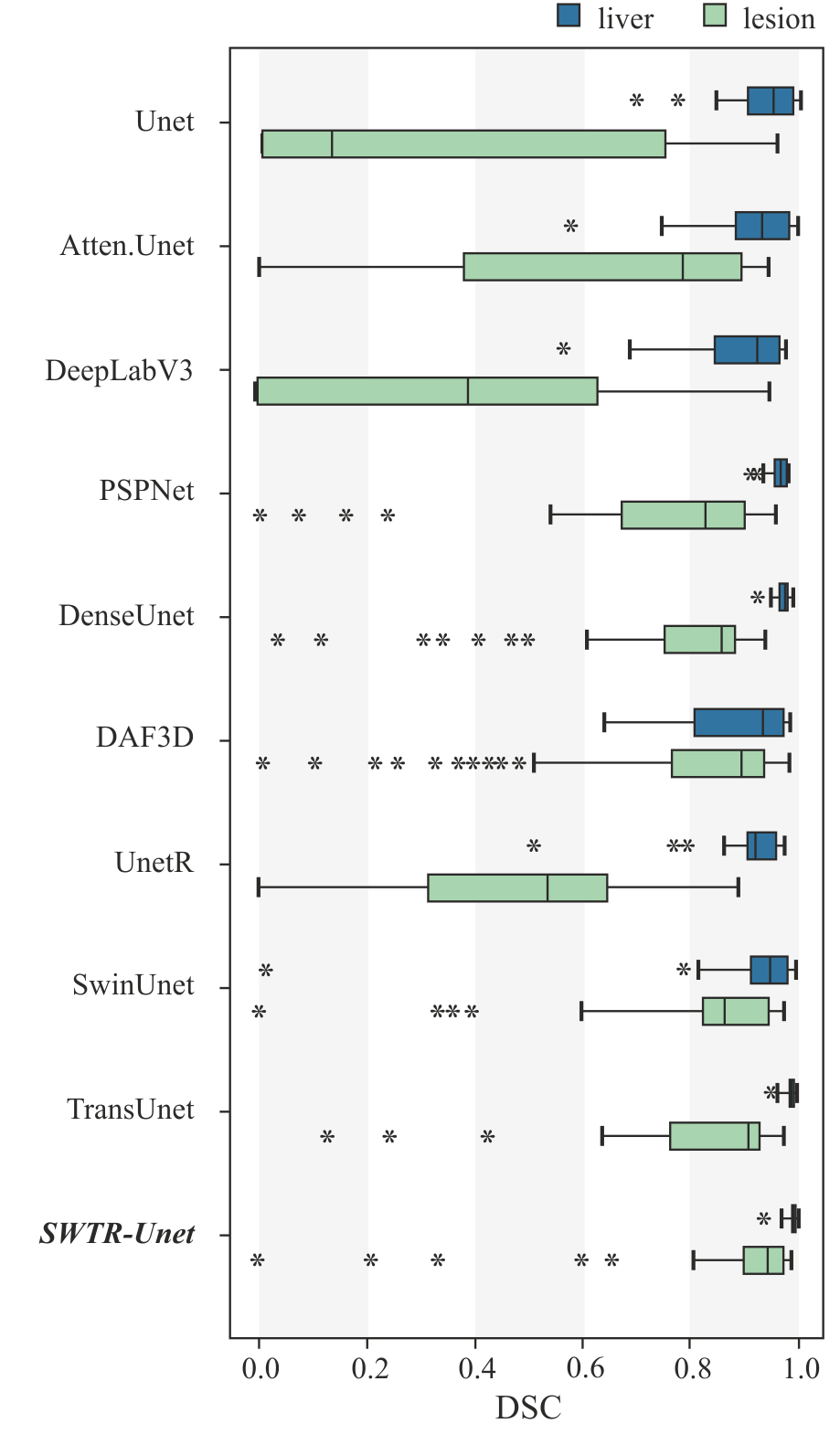}}
\caption{Resulting DSCs averaged over all folds produced by multiple state-of-the-art network variants as well as from the proposed SWTR-Unet. Box edges mark the 25th and 75th percentiles, the central box line marks the median value and the whisker marks the most extreme values not considered as outliers.}
\label{fig1}
\end{figure}

\subsection{MRI-based evaluation}

\begin{figure}[!t]
\centerline{\includegraphics[width=\columnwidth]{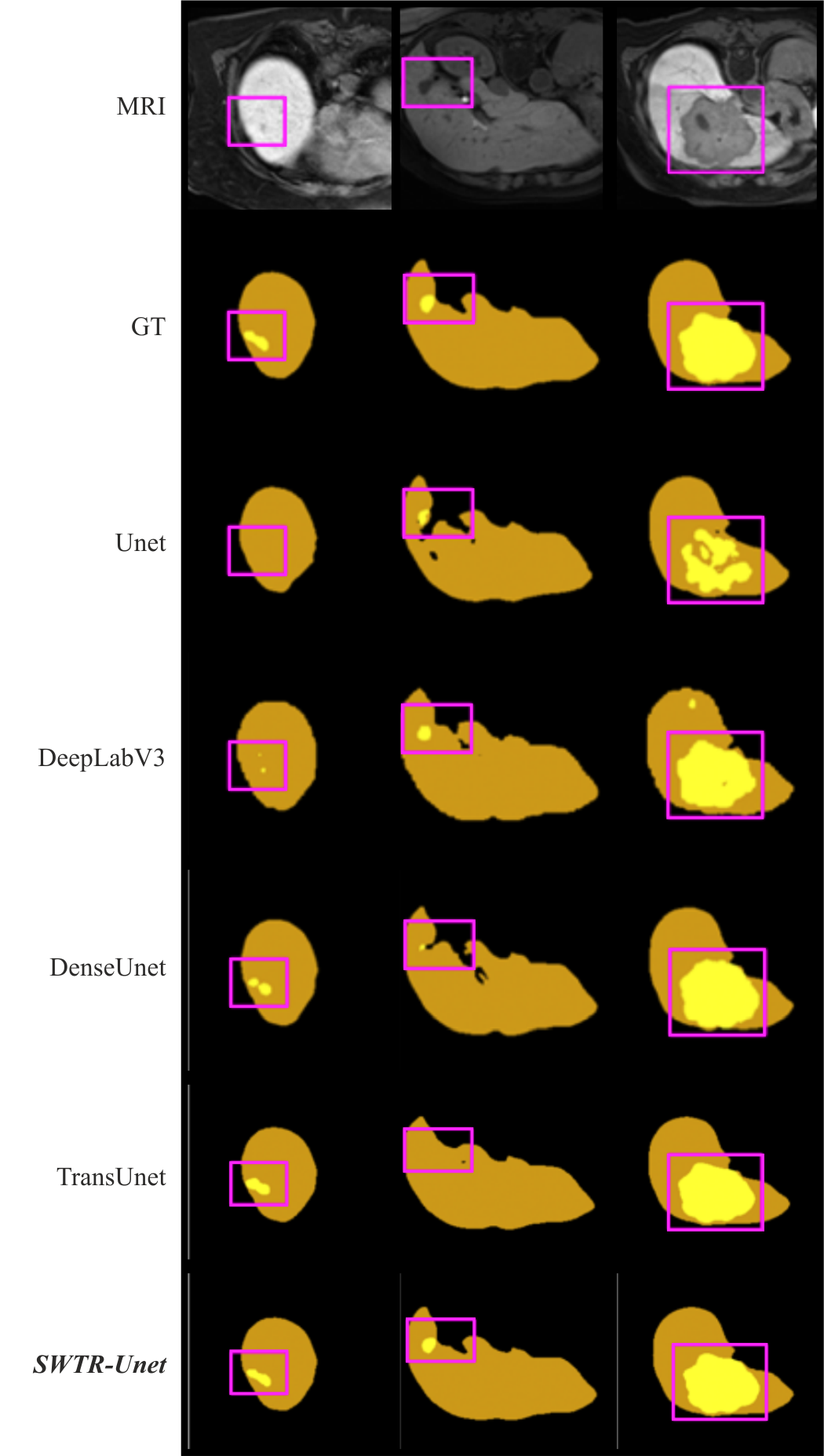}}
\caption{Shown are three exemplary patient cases (from left to right) of the MRI data with their corresponding ground truth (GT) and the networks’ prediction for the liver and lesion mask of various state-of-the-art methods as well as the presented SWTR-Unet (from top to bottom).}
\label{cases}
\end{figure}

Table \ref{tab:mri} shows the averaged segmentation accuracy of all implemented network variants, including both, multiple state-of-the-art architectures and the proposed SWTR-Unet regarding the liver outline and the hepatic lesions therein. With respect to the entirely convolutional-based network variants, the best results could be achieved with the nnUnet with on average $97\%$ liver and $80\%$ lesion Dice. Most of the other state-of-the-art CNNs show promising results regarding the liver segmentation, but lack accuracy in terms of the much more challenging segmentation of the hepatic lesions (see Figure \ref{cases}). The observation of some of the networks producing the worst results revealed several possible reasons for the weak performance: a prevalent cause seems to be false-positive classified pixels in addition to otherwise reasonably good delineated tumors, which occurs on better performing samples as well but is less pronounced there. This problem is most prominent in the case of Unet and DeepLabV3 (average false-positive rate of $41\%$ and $35\%$ as compared to the $8\%$ of the DAF3D network). Furthermore, these networks are not able to sufficiently segment tumors of irregular shapes. Examples are shown in Figure \ref{cases}, where the networks' limitations of segmenting non-circular shaped lesions are demonstrated. Similar behaviour is observed for tumors of small size, which were not appropriately detected by any of the Unet, Attention Unet, and DeepLabV3 networks. Another challenge for those variants represented tumors located close to the liver boundary. In conclusion, the tested convolutional-based networks achieved acceptable segmentation accuracies regarding hepatic lesions of spherical shape, larger size and location in central regions of the liver, but the accuracies drastically decreased for non-spherical shaped and small tumors that are located close to the liver surface.
In contrast, transformer-based networks achieved overall higher segmentation accuracies especially regarding the hepatic lesions. The only exception to this was the UnetR variant, which lacked sufficient accuracy for both object classes. The proposed SWTR-Unet variant outperformed any other network in this test in terms of liver and hepatic lesion segmentation, resulting in DSCs on average of $98\%$ and $81\%$, respectively. Furthermore, in comparison to most of the state-of-the-art network variants, the proposed SWTR-Unet seemed to widely overcome limitations due to smaller sized and non-spherical shaped lesions.

\subsection{CT-based evaluation}

In addition to the MRI-based evaluation, the proposed SWTR-Unet was also applied to the LiTS challenge CT data to verify the applicability on another modality. On this dataset the proposed network achieved DSCs on average of $97 \pm 2\%$ for the liver and $79 \pm 25\%$ for the hepatic lesions therein. The mean Hausdorff distances were $2.04 \pm 2.30~mm$ (liver) and $2.44 \pm 6.30~mm$ (lesion).   

\subsection{Cross-modality evaluation}

In addition to the experiments with solely MRI or CT data, a joint network was designed, which used both data sets as an input (i.e., overall 179 patient cases, including 48 MRI and 131 CT volumes). Analogous to the previous experiments, seven-fold-cross-validation was performed, which resulted in an average accuracy of $0.97 \pm 0.09~\%$ DSC for the liver and $0.78 \pm 0.29~\%$ DSC for the hepatic lesion segmentation. Therefore, the results are slightly inferior compared to both single source experiments, although not far off. This further proves the robustness of the proposed network with respect to the input image modality.

\subsection{Ablation study}

In order to gain more detailed insights of the performance and corresponding influencing factors of the proposed SWTR-Unet on the MRI data set, a number of additional experiments as well as corresponding significance tests (significance level $\alpha < 0.05$) were carried out. 
Starting with varying the number of skip connections (see Table \ref{tab:skip}), which were removed starting with the bottom-most connection (lowest resolution) up to the top-most connection (highest resolution). The results indicate that the skip connections are highly beneficial for the segmentation accuracy and accordingly the accuracy decreases significantly with each removed connection. Therefore, skip connections directly affect the ability of the network to capture lost spatial information during downsampling and thus, higher segmentation accuracies could be achieved.

\begin{table}[htbp]
  \centering
  \caption{Segmentation accuracy of the SWTR-Unet depending on the number of skip connections ($\#_{sc}$). Stated are the average DSCs for liver and hepatic lesions.}

\begin{tabular}{ccc}

\hline
$\#_{sc}$ & $DSC_{liver}$ & $DSC_{lesion}$ \\
\hline
0 & $0.89~\pm~0.08$ & $0.75~\pm~0.15$ \\
1 & $0.91~\pm~0.05$ & $0.77~\pm~0.19$ \\
2 & $0.95~\pm~0.03$ & $0.80~\pm~0.22$ \\
3 & $0.98~\pm~0.03$ & $0.81~\pm~0.27$ \\

\hline   
\end{tabular}%
\label{tab:skip}
\end{table}

Table \ref{tab:trans} shows that extending the network with additional transformer layers increases the DSCs for both liver and tumor segmentation. Transformer layers proved to be in general beneficial to the segmentation accuracy, since they are suitable for capturing long-range dependencies between pixels and thus, global context information. That means, extending the transformer depth to a certain amount of layer, will likely increase the resulting segmentation accuracy, although limited by the drawback of rising computational costs. With the extension from eight to ten and twelve transformer layers the segmentation accuracy could be increased significantly step by step. Only the extension from ten to twelve transformer layers for the liver segmentation proved not to be significant.

\begin{table}[htbp]
  \centering
  \caption{Segmentation accuracy of the SWTR-Unet depending on the number of used transformer layers ($\#_{tl}$). Stated are the average DSCs for liver and hepatic lesions.}

\begin{tabular}{ccc}

\hline
$\#_{tl}$ & $DSC_{liver}$ & $DSC_{lesion}$ \\
\hline
8 & $0.94~\pm~0.06$ & $0.77~\pm~0.17$ \\
10 & $0.97~\pm~0.03$ & $0.79~\pm~0.14$ \\
12 & $0.98~\pm~0.02$ & $0.81~\pm~0.28$ \\

\hline   
\end{tabular}%
\label{tab:trans}
\end{table}

Furthermore, the impact of the number of training samples on the segmentation quality was examined. This was done by training the network with 25, 30, 35, and 45 randomly selected patient cases and their corresponding 2D slices. As illustrated in Table \ref{tab:cases}, an increasing number of training samples led to an overall significantly more precise segmentation result. That should not come as a surprise, since a larger set of independent training data becomes more and more capable to represent the vast variety of the hepatic lesions' size, shape, and location within the liver and therefore, minimizing overfitting behaviour and leading to an improved generalization capability of the network. 

\begin{table}[htbp]
  \centering
  \caption{Segmentation accuracy of the SWTR-Unet depending on the number of used patient cases ($\#_{pc}$) for training purposes. Stated are the average DSCs for liver and hepatic lesions.}

\begin{tabular}{ccc}

\hline
$\#_{pc}$ & $DSC_{liver}$ & $DSC_{lesion}$ \\
\hline
25 & $0.78~\pm~0.10$ & $0.60~\pm~0.18$ \\
30 & $0.86~\pm~0.07$ & $0.72~\pm~0.21$ \\
35 & $0.92~\pm~0.04$ & $0.76~\pm~0.19$ \\
40 & $0.98~\pm~0.02$ & $0.80~\pm~0.21$ \\

\hline   
\end{tabular}%
\label{tab:cases}
\end{table}

Finally, the impact of the pretrained ResNet as well as the presence of the transformer block were investigated (see Table \ref{tab:design}). The results showed a rather marginal benefit of pretraining the ResNet compared to an untrained alternative. This finding thus calls into question the necessity of pretraining the ResNet as a pre-processing step. In contrast, the impact of the transformer bottleneck has proven to be critical, especially with regard to the hepatic lesions, which vastly benefited from its presence. 

\begin{table}[htbp]
  \centering
  \caption{Segmentation accuracy of the SWTR-Unet depending on the usage of a pre- or untrained ResNet within the encoder path, as well as the presence of a transformer-based bottleneck. Stated are the average DSCs for liver and hepatic lesions.}

\begin{tabular}{ccc}

\hline
$ $ & $DSC_{liver}$ & $DSC_{lesion}$ \\
\hline
pretrained ResNet & $0.98~\pm~0.02$ & $0.81~\pm~0.28$ \\
untrained-ResNet & $0.97~\pm~0.03$ & $0.80~\pm~0.29$ \\
\hline
with transformer bottleneck & $0.98~\pm~0.02$ & $0.81~\pm~0.28$ \\
without transformer bottleneck & $0.93~\pm~0.06$ & $0.58~\pm~0.32$ \\

\hline   
\end{tabular}%
\label{tab:design}
\end{table}

\begin{figure}[!t]
\centerline{\includegraphics[width=\columnwidth]{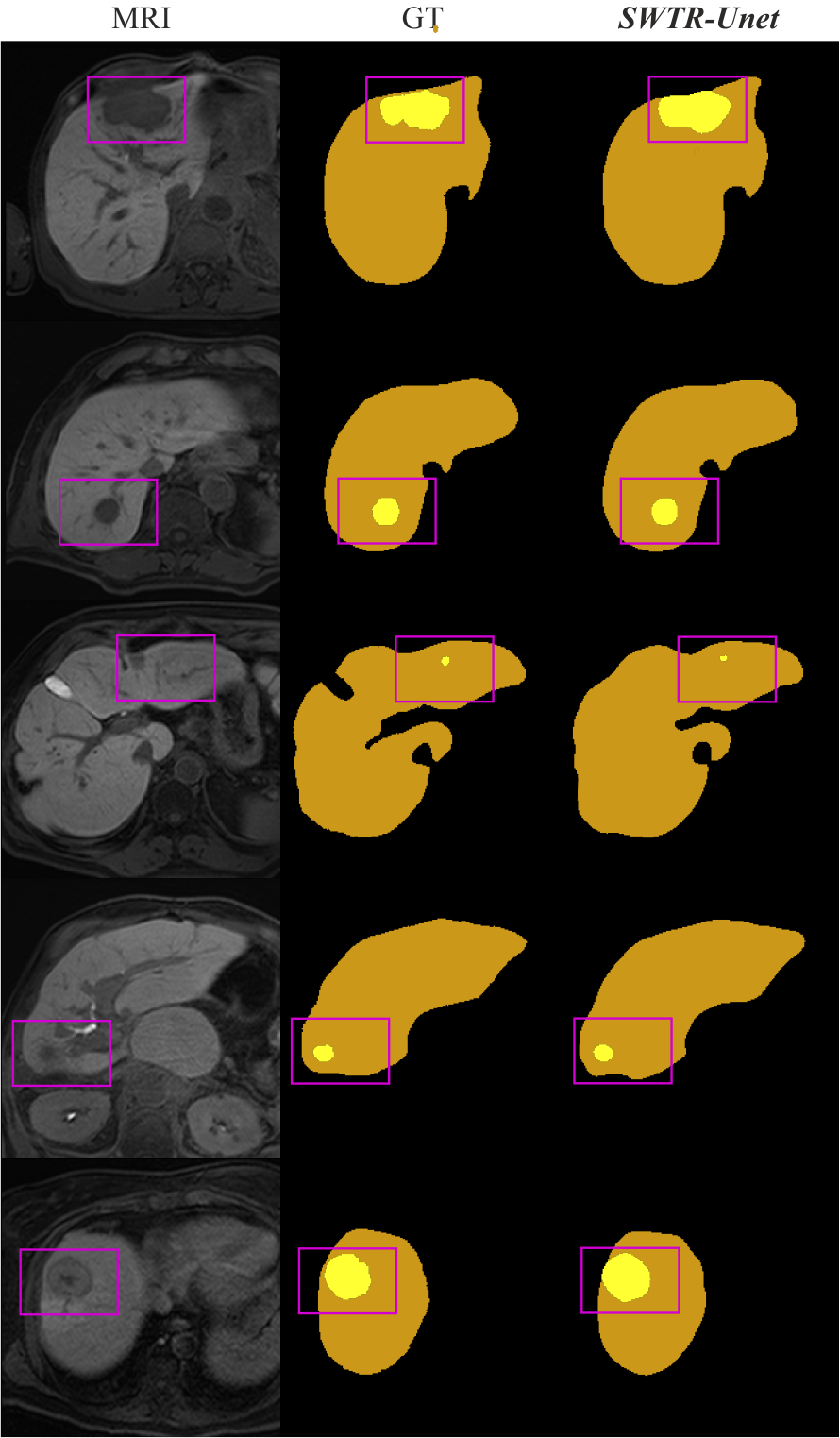}}
\caption{
Shown are five exemplary patient cases (from top to bottom) of the MRI data with their corresponding ground truth (GT) and the network prediction from the presented SWTR-Unet. The first two 
cases display differences in the lesions' sphericity (case 1: $\Psi < 0.9$ with a DSC of $78.1~\%$, 
case 2: $\Psi > 0.9$ with a DSC of $95.9~\%$) and the following three cases highlight the effect of lesion sizes (case 3: $V < 1~cm^3$ with a DSC of $64.9~\%$, case 4: $1< V < 5~cm^3$ with a DSC of $92.1~\%$, case 5: $V > 10~cm^3$ with a DSC of $94.1~\%$).
}
\label{corr}
\end{figure}

\subsection{Correlation analysis}

Due to the high variety of shapes, sizes and locations of the lesions within the liver, it was of utmost interest to examine how these lesion characteristics affect the segmentation accuracy of the proposed SWTR-Unet and to reveal possible bias. This correlation analysis was performed on both, the MRI as well as on the publicly available LiTS challenge CT data (see Table \ref{tab:corr}). \

In terms of lesion shape, the ground truth segmentation of all samples was used to determine the lesions' sphericity $\Psi$ by calculating

\begin{equation}\Psi=\frac{\pi^{\frac{1}{3}} (6V)^{\frac{2}{3}}} {A}, \label{sphere}\end{equation} 

with V being the volume and A the surface of the lesion mask. Subsequently, the lesions were divided into two classes, labelling them as spherical ($\Psi > 0.9$, MRI: 20 lesions, CT: 44 lesions) or merely non-spherically shaped ($\Psi < 0.9$, MRI: 137 lesions, CT: 838). The experiments clearly indicated higher segmentation accuracies if the lesions were of a rather spherical shape with on average $89 \pm 8\%$ DSC (CT: $88 \pm 10\%$ DSC) compared to the class of non-spherically shaped lesions with a mean Dice of $72 \pm 20\%$ DSC (CT: $70 \pm 28\%$ DSC), although these were much more prominently represented in the training set. The prevalence of mere spherically shaped lesions not only increase the chances of more accurate segmentations, but benefits also the robustness of the approach as indicated by the considerably lower standard deviation compared to the results of 
lesions with a $\Psi < 0.9$. \

The volume $V$ of each lesion derived from the ground truth segmentation was used furthermore to divide the whole data set into four different lesion size subgroups: lesions smaller than 1~$cm^3$, between 1 and 5~$cm^3$, between 5 and 10~$cm^3$ and lesions larger than 10~$cm^3$. The results of the experiments using both, the MRI and CT data indicate that the averaged segmentation accuracy as well as the robustness of the method increases with larger tumor sizes (see Table \ref{tab:corr}), which was a similar finding to \citep{b49}. This tendency is further strengthened by the fact that the vastly overrepresented subgroup of the smallest lesions (under 1~$cm^3$, MRI: 68 lesions, CT: 536 lesions) performed statistically significantly worse than the subgroup of the largest lesions (over 10~$cm^3$, MRI: 12 lesions, CT: 117 lesions), despite the fact that the latter tended to be penalised by the sample distribution within the data set.  
 
Finally, the impact of the lesions' inner hepatic location was examined depending on the distance towards the liver surface. Each lesion was labelled based on the distance $d$, that is the closest Euclidean distance between the liver surface and the lesion outline. Hence, it was differentiated between lesions that were surface-near ($d < 1~cm$) or rather centred within the liver ($d > 1~cm$). The MRI results indicate that hepatic lesions close to the liver surface are significantly harder to accurately segment as those, which are more centred within the organ, which is reflected in a remarkably lower mean DSC of $74 \pm 15\%$ compared to on average $87 \pm 5\%$ for rather centred lesion. This may be caused by the more challenging contrast and texture variety at the liver surface with surrounding other tissues, even though both classes were not equally represented within the data set (35 surface-near vs. 122 rather centred cases). The found indication held also true for the segmentations produced on the LiTS challenge data, where lesions at the liver surface scored significantly worse ($70 \pm 11\%$ DSC, 279 lesions) than those in a mere centred location ($86 \pm 6\%$ DSC, 603 lesions). 
  
The impact of each of the investigated influencing factors on the segmentation accuracy was assessed by significance tests (significance level $\alpha < 0.05$). Both the sphericity and the lesion size (smallest class versus largest class), as well as the proximity to the liver surface of the lesions have a statistically significant effect on the segmentation accuracy.

\begin{table}[htbp]
  \centering
  \caption{Results of the correlation analysis depending on the lesions' sphericity $\Psi$, volume $V$ and surface proximity $d$. $\#_{MRI}$ indicates the number of accordingly labelled lesions of the MRI data and  $\#_{CT}$ of the CT data, respectively. Stated are the average DSCs for each class and imaging modality.}

\begin{tabular}{ccccc}

\hline
$ $ & $DSC_{MRI}$ & $\#_{MRI}$ & $DSC_{CT}$ & $\#_{CT}$ \\
\hline
$\Psi<0.9$ & $0.72~\pm~0.20$ & 137 & $0.70~\pm~0.28$ & 838 \\
$\Psi>0.9$ & $0.89~\pm~0.08$ & 20 & $0.88~\pm~0.10$ & 44\\
\hline
$V<1~cm^3$ & $0.78~\pm~0.08$ & 68 & $0.77~\pm~0.09$ & 536 \\
$1<V<5~cm^3$ & $0.80~\pm~0.03$ & 60 & $0.78~\pm~0.08$ & 165 \\
$5<V<10~cm^3$ & $0.82~\pm~0.03$ & 17 & $0.82~\pm~0.06$ & 64 \\
$V>10~cm^3$ & $0.83~\pm~0.01$ & 12 & $0.84~\pm~0.03$ & 117 \\
\hline
$d<1~cm$ & $0.74~\pm~0.15$ & 35 & $0.70~\pm~0.11$ & 279 \\
$d>1~cm$ & $0.87~\pm~0.05$ & 122 & $0.86~\pm~0.06$ & 603 \\

\hline   
\end{tabular}%
\label{tab:corr}
\end{table}

\section{Discussion}

\subsection{MRI-based liver and hepatic lesion segmentation}

Among the first to apply deep learning-based networks to liver lesion segmentation in MRI were \cite{b6}, which included it in their otherwise CT-based study. The authors utilized an Unet-style fully convolutional cascaded neural network with a 3D Conditional Random Field (CRF) for the segmentation of the liver and subsequently using the resulting liver mask as an input for the following lesion segmentation. Regarding the MRI part of their work, they trained their network with 38 patient cases and achieved DSCs on average of $87\%$ for the liver and $69.7\%$ for the hepatic lesion segmentation. \cite{b48} applied an U-net-type network to multiphasic T1-weighted MR images of 174 patient cases to detect and delineate HCC. In order to enhance segmentation accuracy \cite{b48} combined the U-net with a random forest classifier and cluster thresholding and thus, achieved on average DSCs of $91\%$ for liver and $68\%$ for hepatic lesion segmentation. Similar to \cite{b48}, \cite{b49} presented an approach for joint HCC detection and segmentation. However, they shifted away from a more typical U-net-based approach towards a combination of a channel-wise multi-modal image encoder that focusses on complementary multi-modality feature extraction, coordinate sharing between the detection and segmentation path and a subsequent radiomics guided discriminator. \cite{b49} applied their approach to multi-modal (T1-, T2-weighted, DWI) and multiphasic non-contrast-enhanced MR imaging of 255 patient cases. Exploiting the full potential of all MRI sequences and multiphase contrast-enhanced MR imaging they achieved on average a DSC of $83.6 \pm 2.2\%$, whereas single-modality non-contrast-enhanced T1-weighted image input comparable to the image data of this work results in an averaged DSC of $81.1\%$. The most recent work of \cite{b50} utilized a 3D anisotropic U-net and assessed it on late hepatocellular phase contrast-enhanced MR images of 19 test patient cases out of 107 subjects. They matched the resulting segmentation masks of their network with ground truths of three different raters and achieved a DSC of $74 \pm 19\%$ averaged over all test cases and raters.

\begin{table}[t]
  \centering
  \caption{Experimental results of the proposed SWTR-Unet produced on in-house MRI data in comparison with state-of-the-art works. Stated are the Dice similarity coefficients (DSC) of the liver and liver lesion segmentations.}
 \setlength{\tabcolsep}{7pt}
\begin{tabular}{ccc}

\hline
& $DSC_{liver}$ & $DSC_{lesion}$\\
\hline
\cite{b6} & $0.87$ & $0.70$ \\
\cite{b48} & $0.91 \pm 0.01$ & $0.68 \pm 0.03$ \\
\cite{b49} &  & $0.81 \pm 0.03$ \\
\cite{b50} &  & $0.74 \pm 0.19$ \\
\textbf{SWTR-Unet} & $\mathbf{0.98 \pm 0.02}$ & $\mathbf{0.81 \pm 0.28}$ \\
\hline
\end{tabular}%
\label{tab:ct}
\end{table}

Compared to the proposed SWTR-Unet those works, as well as the additionally re-implemented approaches mostly yielded inferior results for hepatic lesion segmentation, especially if they utilized solely convolutional-based architectures, which may not be able to adequately capture the huge shape, size and location variety of hepatic lesions. The addition of non-convolutional elements, e.g., transformer-based building blocks as introduced by \cite{b16}, \cite{b26} and integrated in the proposed SWTR-Unet could significantly enhance the segmentation performance. Another strategy represents the exploitation of complementary multimodal and multiphasic contrast-enhanced MR image features as well as radiomics guided refinements as proposed by \cite{b49}, which yielded the best results so far. Regarding single-modality   
and non-contrast-enhanced MR imaging the proposed SWTR-Unet represents the state of the art on par with the work of \cite{b49}, although in contrast to their work, the SWTR-Unet segments the liver in parallel.
Overall, it is important to note, that with respect to the different data bases, such comparisons are of an indirect nature. In terms of the IOV of MRI-based hepatic lesion segmentation of $78 \pm 12\%$)\cite{b50}, only the proposed SWTR-Unet, the re-implemented nnUnet \citep{b35} and the approach of \cite{b49} could achieve expert level segmentation accuracy.

\subsection{CT-based liver and hepatic lesion segmentation}

In addition to the MRI-based experiments, the SWTR-Unet was applied unmodified to CT data to assess its applicability to another commonly used liver imaging modality. Since most of the works dealing with deep learning-based segmentation of hepatic lesions in CT imaging used the LiTS challenge data, a mere direct comparison of the results is possible (see Table \ref{tab:ct}). 
The segmentation accuracy of the SWTR-Unet proved to be superior to many state-of-the-art works for both liver and hepatic lesion segmentation, except for the works of \cite{b51} and \cite{b53}. The latter utilizes a two-step approach with a preceding detection of the liver lesion bounding box, which is subsequently used as an input for the Unet-based segmentation network. Therefore, the final segmentation accuracy heavily relies on the preceding detection step, which is also reflected by the drop of performance from $83\%$ to $78\%$ if the initial step is omitted. In contrast, the SWTR-Unet works on uncropped, original CT image volumes and additionally delineates the liver. 
\cite{b51} modified an Unet-style architecture by adding a residual path with convolutional layers and activation functions to the skip connections in order to avoid the duplication of low resolution information and enhance higher level feature extraction. In comparison with the SWTR-Unet the work of \cite{b51} achieves similar results regarding the liver segmentation, but proved to be superior with respect to the hepatic lesion segmentation. Both approaches underline the importance of capturing high level features and obtaining global information, which is a common limitation of conventional Unet architectures.

\begin{table}[t]
  \centering
  \caption{Experimental results of the proposed SWTR-Unet (produced on the LiTS challenge CT data) in comparison with state-of-the-art works. Stated are the Dice similarity coefficients (DSC) of the liver and liver lesion segmentations. References marked with an asterisk indicate approaches, that were re-implemented for this work, all other stated results refer to the respective publications.}
 \setlength{\tabcolsep}{7pt}
\begin{tabular}{ccc}

\hline
& $DSC_{liver}$ & $DSC_{lesion}$\\
\hline
\cite{b11}      & $0.97 $ & $0.69$ \\
\cite{b9}    & $0.96 $ & $0.68$ \\
\cite{b25} & $0.95 $ & $0.66$ \\
\cite{b13}       & $0.96 \pm 0.03$ & $0.74 \pm 0.08$ \\
\cite{b51} & $0.99 \pm 0.01$ & $0.90 \pm 0.05$ \\
\cite{b52} & $0.96 \pm 0.05$ & $0.76 \pm 0.03$ \\
\cite{b53} &  & $0.83$ \\
\cite{b35}*   & $0.97 \pm 0.04$ & $0.76 \pm 0.23$ \\
\cite{b26}* & $0.93 \pm 0.06$ & $0.76 \pm 0.27$ \\
\cite{b16}* & $0.96 \pm 0.02$ & $0.77 \pm 0.21$ \\
\textbf{SWTR-Unet} & $\mathbf{0.98 \pm 0.02}$ & $\mathbf{0.79 \pm 0.25}$ \\
\hline
\end{tabular}%
\label{tab:ct}
\end{table}

\subsection{Conclusion}

Precise segmentations of the liver and hepatic tumors are of utmost importance since decisions regarding the proper treatment evermore depend on findings provided by such procedures. Whereas manual segmentation represents the gold standard in terms of accuracy, it is time-consuming, cumbersome, and unnecessarily ties up valuable resources, which is why automatized procedures gain relevance in clinical settings. 

In this work, we presented a novel hybrid network architecture combining convolutional and transformer-based elements and compared it with additionally implemented state-of-the-art approaches on the same evaluation data. This ensures a direct comparability of all methods, which is otherwise most often limited due to different databases of the related work. In this regard, all network variants were applied to clinical MRI data of the University Hospital of Magdeburg. Furthermore, the proposed approach was applied to publicly available CT imaging data of the LiTS challenge. In order to investigate various influencing factors on the segmentation accuracy, various parameters of the network architecture, as well as the influence of lesion size, shape and location on the results were examined.  

Based on these experiments, the proposed SWTR-Unet achieved highly promising segmentation accuracies regarding both, the liver and its hepatic lesions with DSCs on average of $98 \pm 3\%$ and $81 \pm 25\%$ on MRI and $98 \pm 2\%$ and $79 \pm 25\%$ on CT data. In comparison to the related work, the proposed approach represents the current state of the art in MRI-based hepatic lesion segmentation. The findings of further experiments underlined the impact of lesion size, shape and location within the liver, which indicated that the segmentation accuracy increases with larger, more spherical and rather centred lesions. Furthermore, it could be shown that an increasing number of training samples, skip connections and transformer layers have a beneficial effect on the segmentation accuracy.
In conclusion, the proposed SWTR-Unet could represent an important step towards a more sophisticated computer-assisted workflow of liver lesion diagnosis and therapy by providing expert-level segmentation accuracy with little to no required user interaction. Therefore, it could support the radiologists in clinical practice by saving valuable resources and time.\\

\section{Acknowledgements}

\noindent \textbf{Funding:} The work of this paper was funded by the Federal Ministry of Education and Research within the  Forschungscampus STIMULATE under grant number '13GW0473A’.

\noindent \textbf{Ethical approval:} All procedures performed in studies involving human
participants were in accordance with the ethical standards of the institutional
and/or national research committee and with the 1964 Helsinki declaration
and its later amendments or comparable ethical standards. For this
type of study formal consent is not required.

\noindent \textbf{Conflict of Interest Statement:} None declared.

%%Harvard
\bibliographystyle{model2-names.bst}\biboptions{authoryear}
\bibliography{arxiv_update}

\begin{thebibliography}{43}
\expandafter\ifx\csname natexlab\endcsname\relax\def\natexlab#1{#1}\fi
\providecommand{\url}[1]{\texttt{#1}}
\providecommand{\href}[2]{#2}
\providecommand{\path}[1]{#1}
\providecommand{\DOIprefix}{doi:}
\providecommand{\ArXivprefix}{arXiv:}
\providecommand{\URLprefix}{URL: }
\providecommand{\Pubmedprefix}{pmid:}
\providecommand{\doi}[1]{\href{http://dx.doi.org/#1}{\path{#1}}}
\providecommand{\Pubmed}[1]{\href{pmid:#1}{\path{#1}}}
\providecommand{\bibinfo}[2]{#2}
\ifx\xfnm\relax \def\xfnm[#1]{\unskip,\space#1}\fi
%Type = Article
\bibitem[{Ara{\'u}jo et~al.(2021)Ara{\'u}jo, da~Cruz, Ferreira, da~Silva~Neto,
  Silva, de~Paiva and Gattass}]{b53}
\bibinfo{author}{Ara{\'u}jo, J.D.L.}, \bibinfo{author}{da~Cruz, L.B.},
  \bibinfo{author}{Ferreira, J.L.}, \bibinfo{author}{da~Silva~Neto, O.P.},
  \bibinfo{author}{Silva, A.C.}, \bibinfo{author}{de~Paiva, A.C.},
  \bibinfo{author}{Gattass, M.}, \bibinfo{year}{2021}.
\newblock \bibinfo{title}{An automatic method for segmentation of liver lesions
  in computed tomography images using deep neural networks}.
\newblock \bibinfo{journal}{Expert Systems with Applications}
  \bibinfo{volume}{180}, \bibinfo{pages}{115064}.
%Type = Article
\bibitem[{Bilic et~al.(2019)Bilic, Christ, Vorontsov, Chlebus, Chen, Dou, Fu,
  Han, Heng, Hesser et~al.}]{b20}
\bibinfo{author}{Bilic, P.}, \bibinfo{author}{Christ, P.F.},
  \bibinfo{author}{Vorontsov, E.}, \bibinfo{author}{Chlebus, G.},
  \bibinfo{author}{Chen, H.}, \bibinfo{author}{Dou, Q.}, \bibinfo{author}{Fu,
  C.W.}, \bibinfo{author}{Han, X.}, \bibinfo{author}{Heng, P.A.},
  \bibinfo{author}{Hesser, J.}, et~al., \bibinfo{year}{2019}.
\newblock \bibinfo{title}{The liver tumor segmentation benchmark (lits)}.
\newblock \bibinfo{journal}{arXiv preprint arXiv:1901.04056} .
%Type = Article
\bibitem[{Bousabarah et~al.(2021)Bousabarah, Letzen, Tefera, Savic, Schobert,
  Schlachter, Staib, Kocher, Chapiro and Lin}]{b48}
\bibinfo{author}{Bousabarah, K.}, \bibinfo{author}{Letzen, B.},
  \bibinfo{author}{Tefera, J.}, \bibinfo{author}{Savic, L.},
  \bibinfo{author}{Schobert, I.}, \bibinfo{author}{Schlachter, T.},
  \bibinfo{author}{Staib, L.H.}, \bibinfo{author}{Kocher, M.},
  \bibinfo{author}{Chapiro, J.}, \bibinfo{author}{Lin, M.},
  \bibinfo{year}{2021}.
\newblock \bibinfo{title}{Automated detection and delineation of hepatocellular
  carcinoma on multiphasic contrast-enhanced mri using deep learning}.
\newblock \bibinfo{journal}{Abdominal Radiology} \bibinfo{volume}{46},
  \bibinfo{pages}{216--225}.
%Type = Article
\bibitem[{Cai et~al.(2020)Cai, Tian, Lui, Zeng, Wu and Chen}]{b32}
\bibinfo{author}{Cai, S.}, \bibinfo{author}{Tian, Y.}, \bibinfo{author}{Lui,
  H.}, \bibinfo{author}{Zeng, H.}, \bibinfo{author}{Wu, Y.},
  \bibinfo{author}{Chen, G.}, \bibinfo{year}{2020}.
\newblock \bibinfo{title}{Dense-unet: a novel multiphoton in vivo cellular
  image segmentation model based on a convolutional neural network}.
\newblock \bibinfo{journal}{Quantitative imaging in medicine and surgery}
  \bibinfo{volume}{10}, \bibinfo{pages}{1275}.
%Type = Article
\bibitem[{Cao et~al.(2021)Cao, Wang, Chen, Jiang, Zhang, Tian and Wang}]{b26}
\bibinfo{author}{Cao, H.}, \bibinfo{author}{Wang, Y.}, \bibinfo{author}{Chen,
  J.}, \bibinfo{author}{Jiang, D.}, \bibinfo{author}{Zhang, X.},
  \bibinfo{author}{Tian, Q.}, \bibinfo{author}{Wang, M.}, \bibinfo{year}{2021}.
\newblock \bibinfo{title}{Swin-unet: Unet-like pure transformer for medical
  image segmentation}.
\newblock \bibinfo{journal}{arXiv preprint arXiv:2105.05537} .
%Type = Article
\bibitem[{Chen et~al.(2021)Chen, Lu, Yu, Luo, Adeli, Wang, Lu, Yuille and
  Zhou}]{b16}
\bibinfo{author}{Chen, J.}, \bibinfo{author}{Lu, Y.}, \bibinfo{author}{Yu, Q.},
  \bibinfo{author}{Luo, X.}, \bibinfo{author}{Adeli, E.},
  \bibinfo{author}{Wang, Y.}, \bibinfo{author}{Lu, L.},
  \bibinfo{author}{Yuille, A.L.}, \bibinfo{author}{Zhou, Y.},
  \bibinfo{year}{2021}.
\newblock \bibinfo{title}{Transunet: Transformers make strong encoders for
  medical image segmentation}.
\newblock \bibinfo{journal}{arXiv preprint arXiv:2102.04306} .
%Type = Article
\bibitem[{Chen et~al.(2017)Chen, Papandreou, Schroff and Adam}]{b29}
\bibinfo{author}{Chen, L.C.}, \bibinfo{author}{Papandreou, G.},
  \bibinfo{author}{Schroff, F.}, \bibinfo{author}{Adam, H.},
  \bibinfo{year}{2017}.
\newblock \bibinfo{title}{Rethinking atrous convolution for semantic image
  segmentation}.
\newblock \bibinfo{journal}{arXiv preprint arXiv:1706.05587} .
%Type = Article
\bibitem[{Chi et~al.(2021)Chi, Han, Wu, Wang and Ji}]{b52}
\bibinfo{author}{Chi, J.}, \bibinfo{author}{Han, X.}, \bibinfo{author}{Wu, C.},
  \bibinfo{author}{Wang, H.}, \bibinfo{author}{Ji, P.}, \bibinfo{year}{2021}.
\newblock \bibinfo{title}{X-net: Multi-branch unet-like network for liver and
  tumor segmentation from 3d abdominal ct scans}.
\newblock \bibinfo{journal}{Neurocomputing} \bibinfo{volume}{459},
  \bibinfo{pages}{81--96}.
%Type = Article
\bibitem[{Child et~al.(2019)Child, Gray, Radford and Sutskever}]{b42}
\bibinfo{author}{Child, R.}, \bibinfo{author}{Gray, S.},
  \bibinfo{author}{Radford, A.}, \bibinfo{author}{Sutskever, I.},
  \bibinfo{year}{2019}.
\newblock \bibinfo{title}{Generating long sequences with sparse transformers}.
\newblock \bibinfo{journal}{arXiv preprint arXiv:1904.10509} .
%Type = Article
\bibitem[{Chlebus et~al.(2018)Chlebus, Schenk, Moltz, van Ginneken, Hahn and
  Meine}]{b9}
\bibinfo{author}{Chlebus, G.}, \bibinfo{author}{Schenk, A.},
  \bibinfo{author}{Moltz, J.H.}, \bibinfo{author}{van Ginneken, B.},
  \bibinfo{author}{Hahn, H.K.}, \bibinfo{author}{Meine, H.},
  \bibinfo{year}{2018}.
\newblock \bibinfo{title}{Automatic liver tumor segmentation in ct with fully
  convolutional neural networks and object-based postprocessing}.
\newblock \bibinfo{journal}{Scientific reports} \bibinfo{volume}{8},
  \bibinfo{pages}{1--7}.
%Type = Article
\bibitem[{Christ et~al.(2017)Christ, Ettlinger, Gr{\"u}n, Elshaera, Lipkova,
  Schlecht, Ahmaddy, Tatavarty, Bickel, Bilic, Rempfler, Hofmann, Anastasi,
  Ahmadi, Kaissis, Holch, Sommer, Braren, Heinemann and Menze}]{b6}
\bibinfo{author}{Christ, P.F.}, \bibinfo{author}{Ettlinger, F.},
  \bibinfo{author}{Gr{\"u}n, F.}, \bibinfo{author}{Elshaera, M.E.A.},
  \bibinfo{author}{Lipkova, J.}, \bibinfo{author}{Schlecht, S.},
  \bibinfo{author}{Ahmaddy, F.}, \bibinfo{author}{Tatavarty, S.},
  \bibinfo{author}{Bickel, M.}, \bibinfo{author}{Bilic, P.},
  \bibinfo{author}{Rempfler, M.}, \bibinfo{author}{Hofmann, F.},
  \bibinfo{author}{Anastasi, M.}, \bibinfo{author}{Ahmadi, S.A.},
  \bibinfo{author}{Kaissis, G.}, \bibinfo{author}{Holch, J.},
  \bibinfo{author}{Sommer, W.}, \bibinfo{author}{Braren, R.},
  \bibinfo{author}{Heinemann, V.}, \bibinfo{author}{Menze, B.},
  \bibinfo{year}{2017}.
\newblock \bibinfo{title}{Automatic liver and tumor segmentation of ct and mri
  volumes using cascaded fully convolutional neural networks}.
\newblock \bibinfo{journal}{arXiv preprint arXiv:1702.05970} .
%Type = Inproceedings
\bibitem[{Deng et~al.(2009)Deng, Dong, Socher, Li, Li and Fei-Fei}]{b15}
\bibinfo{author}{Deng, J.}, \bibinfo{author}{Dong, W.},
  \bibinfo{author}{Socher, R.}, \bibinfo{author}{Li, L.J.},
  \bibinfo{author}{Li, K.}, \bibinfo{author}{Fei-Fei, L.},
  \bibinfo{year}{2009}.
\newblock \bibinfo{title}{Imagenet: A large-scale hierarchical image database},
  in: \bibinfo{booktitle}{2009 IEEE conference on computer vision and pattern
  recognition}, \bibinfo{organization}{Ieee}. pp. \bibinfo{pages}{248--255}.
%Type = Article
\bibitem[{Fan et~al.(2020)Fan, Wang, Li and Wang}]{b13}
\bibinfo{author}{Fan, T.}, \bibinfo{author}{Wang, G.}, \bibinfo{author}{Li,
  Y.}, \bibinfo{author}{Wang, H.}, \bibinfo{year}{2020}.
\newblock \bibinfo{title}{Ma-net: A multi-scale attention network for liver and
  tumor segmentation}.
\newblock \bibinfo{journal}{IEEE Access} \bibinfo{volume}{8},
  \bibinfo{pages}{179656--179665}.
%Type = Article
\bibitem[{Ferlay et~al.(2010)Ferlay, Shin, Bray, Forman, Mathers and
  Parkin}]{b2}
\bibinfo{author}{Ferlay, J.}, \bibinfo{author}{Shin, H.R.},
  \bibinfo{author}{Bray, F.}, \bibinfo{author}{Forman, D.},
  \bibinfo{author}{Mathers, C.}, \bibinfo{author}{Parkin, D.M.},
  \bibinfo{year}{2010}.
\newblock \bibinfo{title}{Estimates of worldwide burden of cancer in 2008:
  Globocan 2008}.
\newblock \bibinfo{journal}{International journal of cancer}
  \bibinfo{volume}{127}, \bibinfo{pages}{2893--2917}.
%Type = Article
\bibitem[{H{\"a}nsch et~al.(2022)H{\"a}nsch, Chlebus, Meine, Thielke, Kock,
  Paulus, Abolmaali and Schenk}]{b50}
\bibinfo{author}{H{\"a}nsch, A.}, \bibinfo{author}{Chlebus, G.},
  \bibinfo{author}{Meine, H.}, \bibinfo{author}{Thielke, F.},
  \bibinfo{author}{Kock, F.}, \bibinfo{author}{Paulus, T.},
  \bibinfo{author}{Abolmaali, N.}, \bibinfo{author}{Schenk, A.},
  \bibinfo{year}{2022}.
\newblock \bibinfo{title}{Improving automatic liver tumor segmentation in
  late-phase mri using multi-model training and 3d convolutional neural
  networks}.
\newblock \bibinfo{journal}{Scientific Reports} \bibinfo{volume}{12},
  \bibinfo{pages}{1--10}.
%Type = Inproceedings
\bibitem[{Hatamizadeh et~al.(2022)Hatamizadeh, Tang, Nath, Yang, Myronenko,
  Landman, Roth and Xu}]{b31}
\bibinfo{author}{Hatamizadeh, A.}, \bibinfo{author}{Tang, Y.},
  \bibinfo{author}{Nath, V.}, \bibinfo{author}{Yang, D.},
  \bibinfo{author}{Myronenko, A.}, \bibinfo{author}{Landman, B.},
  \bibinfo{author}{Roth, H.R.}, \bibinfo{author}{Xu, D.}, \bibinfo{year}{2022}.
\newblock \bibinfo{title}{Unetr: Transformers for 3d medical image
  segmentation}, in: \bibinfo{booktitle}{Proceedings of the IEEE/CVF Winter
  Conference on Applications of Computer Vision}, pp.
  \bibinfo{pages}{574--584}.
%Type = Inproceedings
\bibitem[{He et~al.(2016)He, Zhang, Ren and Sun}]{b14}
\bibinfo{author}{He, K.}, \bibinfo{author}{Zhang, X.}, \bibinfo{author}{Ren,
  S.}, \bibinfo{author}{Sun, J.}, \bibinfo{year}{2016}.
\newblock \bibinfo{title}{Deep residual learning for image recognition}, in:
  \bibinfo{booktitle}{Proceedings of the IEEE conference on computer vision and
  pattern recognition}, pp. \bibinfo{pages}{770--778}.
%Type = Article
\bibitem[{Isensee et~al.(2021)Isensee, Jaeger, Kohl, Petersen and
  Maier-Hein}]{b35}
\bibinfo{author}{Isensee, F.}, \bibinfo{author}{Jaeger, P.F.},
  \bibinfo{author}{Kohl, S.A.}, \bibinfo{author}{Petersen, J.},
  \bibinfo{author}{Maier-Hein, K.H.}, \bibinfo{year}{2021}.
\newblock \bibinfo{title}{nnu-net: a self-configuring method for deep
  learning-based biomedical image segmentation}.
\newblock \bibinfo{journal}{Nature methods} \bibinfo{volume}{18},
  \bibinfo{pages}{203--211}.
%Type = Article
\bibitem[{Li et~al.(2013)Li, Wang, Eberl, Fulham, Yin, Chen and Feng}]{b4}
\bibinfo{author}{Li, C.}, \bibinfo{author}{Wang, X.}, \bibinfo{author}{Eberl,
  S.}, \bibinfo{author}{Fulham, M.}, \bibinfo{author}{Yin, Y.},
  \bibinfo{author}{Chen, J.}, \bibinfo{author}{Feng, D.D.},
  \bibinfo{year}{2013}.
\newblock \bibinfo{title}{A likelihood and local constraint level set model for
  liver tumor segmentation from ct volumes}.
\newblock \bibinfo{journal}{IEEE Transactions on Biomedical Engineering}
  \bibinfo{volume}{60}, \bibinfo{pages}{2967--2977}.
%Type = Article
\bibitem[{Li et~al.(2015)Li, Chen, Shi, Zhu, Tian and Xiang}]{b3}
\bibinfo{author}{Li, G.}, \bibinfo{author}{Chen, X.}, \bibinfo{author}{Shi,
  F.}, \bibinfo{author}{Zhu, W.}, \bibinfo{author}{Tian, J.},
  \bibinfo{author}{Xiang, D.}, \bibinfo{year}{2015}.
\newblock \bibinfo{title}{Automatic liver segmentation based on shape
  constraints and deformable graph cut in ct images}.
\newblock \bibinfo{journal}{IEEE Transactions on Image Processing}
  \bibinfo{volume}{24}, \bibinfo{pages}{5315--5329}.
%Type = Article
\bibitem[{Linguraru et~al.(2012)Linguraru, Richbourg, Liu, Watt, Pamulapati,
  Wang and Summers}]{b5}
\bibinfo{author}{Linguraru, M.G.}, \bibinfo{author}{Richbourg, W.J.},
  \bibinfo{author}{Liu, J.}, \bibinfo{author}{Watt, J.M.},
  \bibinfo{author}{Pamulapati, V.}, \bibinfo{author}{Wang, S.},
  \bibinfo{author}{Summers, R.M.}, \bibinfo{year}{2012}.
\newblock \bibinfo{title}{Tumor burden analysis on computed tomography by
  automated liver and tumor segmentation}.
\newblock \bibinfo{journal}{IEEE transactions on medical imaging}
  \bibinfo{volume}{31}, \bibinfo{pages}{1965--1976}.
%Type = Inproceedings
\bibitem[{Liu et~al.(2021)Liu, Lin, Cao, Hu, Wei, Zhang, Lin and Guo}]{b18}
\bibinfo{author}{Liu, Z.}, \bibinfo{author}{Lin, Y.}, \bibinfo{author}{Cao,
  Y.}, \bibinfo{author}{Hu, H.}, \bibinfo{author}{Wei, Y.},
  \bibinfo{author}{Zhang, Z.}, \bibinfo{author}{Lin, S.}, \bibinfo{author}{Guo,
  B.}, \bibinfo{year}{2021}.
\newblock \bibinfo{title}{Swin transformer: Hierarchical vision transformer
  using shifted windows}, in: \bibinfo{booktitle}{Proceedings of the IEEE/CVF
  International Conference on Computer Vision}, pp.
  \bibinfo{pages}{10012--10022}.
%Type = Article
\bibitem[{Meng et~al.(2020)Meng, Tian and Bu}]{b11}
\bibinfo{author}{Meng, L.}, \bibinfo{author}{Tian, Y.}, \bibinfo{author}{Bu,
  S.}, \bibinfo{year}{2020}.
\newblock \bibinfo{title}{Liver tumor segmentation based on 3d convolutional
  neural network with dual scale}.
\newblock \bibinfo{journal}{Journal of applied clinical medical physics}
  \bibinfo{volume}{21}, \bibinfo{pages}{144--157}.
%Type = Article
\bibitem[{Oktay et~al.(2018)Oktay, Schlemper, Folgoc, Lee, Heinrich, Misawa,
  Mori, McDonagh, Hammerla, Kainz, Glocker and Rueckert}]{b28}
\bibinfo{author}{Oktay, O.}, \bibinfo{author}{Schlemper, J.},
  \bibinfo{author}{Folgoc, L.L.}, \bibinfo{author}{Lee, M.},
  \bibinfo{author}{Heinrich, M.}, \bibinfo{author}{Misawa, K.},
  \bibinfo{author}{Mori, K.}, \bibinfo{author}{McDonagh, S.},
  \bibinfo{author}{Hammerla, N.Y.}, \bibinfo{author}{Kainz, B.},
  \bibinfo{author}{Glocker, B.}, \bibinfo{author}{Rueckert, D.},
  \bibinfo{year}{2018}.
\newblock \bibinfo{title}{Attention u-net: Learning where to look for the
  pancreas}.
\newblock \bibinfo{journal}{arXiv preprint arXiv:1804.03999} .
%Type = Inproceedings
\bibitem[{Parmar et~al.(2018)Parmar, Vaswani, Uszkoreit, Kaiser, Shazeer, Ku
  and Tran}]{b43}
\bibinfo{author}{Parmar, N.}, \bibinfo{author}{Vaswani, A.},
  \bibinfo{author}{Uszkoreit, J.}, \bibinfo{author}{Kaiser, L.},
  \bibinfo{author}{Shazeer, N.}, \bibinfo{author}{Ku, A.},
  \bibinfo{author}{Tran, D.}, \bibinfo{year}{2018}.
\newblock \bibinfo{title}{Image transformer}, in:
  \bibinfo{booktitle}{International conference on machine learning},
  \bibinfo{organization}{PMLR}. pp. \bibinfo{pages}{4055--4064}.
%Type = Incollection
\bibitem[{Poudel et~al.(2016)Poudel, Lamata and Montana}]{b46}
\bibinfo{author}{Poudel, R.P.}, \bibinfo{author}{Lamata, P.},
  \bibinfo{author}{Montana, G.}, \bibinfo{year}{2016}.
\newblock \bibinfo{title}{Recurrent fully convolutional neural networks for
  multi-slice mri cardiac segmentation}, in:
  \bibinfo{booktitle}{Reconstruction, segmentation, and analysis of medical
  images}. \bibinfo{publisher}{Springer}, pp. \bibinfo{pages}{83--94}.
%Type = Article
\bibitem[{Rezaei et~al.(2020)Rezaei, Yang and Meinel}]{b47}
\bibinfo{author}{Rezaei, M.}, \bibinfo{author}{Yang, H.},
  \bibinfo{author}{Meinel, C.}, \bibinfo{year}{2020}.
\newblock \bibinfo{title}{Recurrent generative adversarial network for learning
  imbalanced medical image semantic segmentation}.
\newblock \bibinfo{journal}{Multimedia Tools and Applications}
  \bibinfo{volume}{79}, \bibinfo{pages}{15329--15348}.
%Type = Inproceedings
\bibitem[{Ronneberger et~al.(2015)Ronneberger, Fischer and Brox}]{b27}
\bibinfo{author}{Ronneberger, O.}, \bibinfo{author}{Fischer, P.},
  \bibinfo{author}{Brox, T.}, \bibinfo{year}{2015}.
\newblock \bibinfo{title}{U-net: Convolutional networks for biomedical image
  segmentation}, in: \bibinfo{booktitle}{International Conference on Medical
  image computing and computer-assisted intervention},
  \bibinfo{organization}{Springer}. pp. \bibinfo{pages}{234--241}.
%Type = Article
\bibitem[{Schlemper et~al.(2019a)Schlemper, Oktay, Schaap, Heinrich, Kainz,
  Glocker and Rueckert}]{b37}
\bibinfo{author}{Schlemper, J.}, \bibinfo{author}{Oktay, O.},
  \bibinfo{author}{Schaap, M.}, \bibinfo{author}{Heinrich, M.},
  \bibinfo{author}{Kainz, B.}, \bibinfo{author}{Glocker, B.},
  \bibinfo{author}{Rueckert, D.}, \bibinfo{year}{2019}a.
\newblock \bibinfo{title}{Attention gated networks: Learning to leverage
  salient regions in medical images}.
\newblock \bibinfo{journal}{Medical image analysis} \bibinfo{volume}{53},
  \bibinfo{pages}{197--207}.
%Type = Article
\bibitem[{Schlemper et~al.(2019b)Schlemper, Oktay, Schaap, Heinrich, Kainz,
  Glocker and Rueckert}]{b40}
\bibinfo{author}{Schlemper, J.}, \bibinfo{author}{Oktay, O.},
  \bibinfo{author}{Schaap, M.}, \bibinfo{author}{Heinrich, M.},
  \bibinfo{author}{Kainz, B.}, \bibinfo{author}{Glocker, B.},
  \bibinfo{author}{Rueckert, D.}, \bibinfo{year}{2019}b.
\newblock \bibinfo{title}{Attention gated networks: Learning to leverage
  salient regions in medical images}.
\newblock \bibinfo{journal}{Medical image analysis} \bibinfo{volume}{53},
  \bibinfo{pages}{197--207}.
%Type = Article
\bibitem[{Seo et~al.(2019)Seo, Huang, Bassenne, Xiao and Xing}]{b51}
\bibinfo{author}{Seo, H.}, \bibinfo{author}{Huang, C.},
  \bibinfo{author}{Bassenne, M.}, \bibinfo{author}{Xiao, R.},
  \bibinfo{author}{Xing, L.}, \bibinfo{year}{2019}.
\newblock \bibinfo{title}{Modified u-net (mu-net) with incorporation of
  object-dependent high level features for improved liver and liver-tumor
  segmentation in ct images}.
\newblock \bibinfo{journal}{IEEE transactions on medical imaging}
  \bibinfo{volume}{39}, \bibinfo{pages}{1316--1325}.
%Type = Inproceedings
\bibitem[{Touvron et~al.(2021)Touvron, Cord, Douze, Massa, Sablayrolles and
  J{\'e}gou}]{b44}
\bibinfo{author}{Touvron, H.}, \bibinfo{author}{Cord, M.},
  \bibinfo{author}{Douze, M.}, \bibinfo{author}{Massa, F.},
  \bibinfo{author}{Sablayrolles, A.}, \bibinfo{author}{J{\'e}gou, H.},
  \bibinfo{year}{2021}.
\newblock \bibinfo{title}{Training data-efficient image transformers \&
  distillation through attention}, in: \bibinfo{booktitle}{International
  Conference on Machine Learning}, \bibinfo{organization}{PMLR}. pp.
  \bibinfo{pages}{10347--10357}.
%Type = Inproceedings
\bibitem[{Valanarasu et~al.(2021)Valanarasu, Oza, Hacihaliloglu and
  Patel}]{b41}
\bibinfo{author}{Valanarasu, J.M.J.}, \bibinfo{author}{Oza, P.},
  \bibinfo{author}{Hacihaliloglu, I.}, \bibinfo{author}{Patel, V.M.},
  \bibinfo{year}{2021}.
\newblock \bibinfo{title}{Medical transformer: Gated axial-attention for
  medical image segmentation}, in: \bibinfo{booktitle}{International Conference
  on Medical Image Computing and Computer-Assisted Intervention},
  \bibinfo{organization}{Springer}. pp. \bibinfo{pages}{36--46}.
%Type = Article
\bibitem[{Vaswani et~al.(2017)Vaswani, Shazeer, Parmar, Uszkoreit, Jones,
  Gomez, Kaiser and Polosukhin}]{b17}
\bibinfo{author}{Vaswani, A.}, \bibinfo{author}{Shazeer, N.},
  \bibinfo{author}{Parmar, N.}, \bibinfo{author}{Uszkoreit, J.},
  \bibinfo{author}{Jones, L.}, \bibinfo{author}{Gomez, A.N.},
  \bibinfo{author}{Kaiser, {\L}.}, \bibinfo{author}{Polosukhin, I.},
  \bibinfo{year}{2017}.
\newblock \bibinfo{title}{Attention is all you need}.
\newblock \bibinfo{journal}{Advances in neural information processing systems}
  \bibinfo{volume}{30}.
%Type = Article
\bibitem[{Vorontsov et~al.(2019)Vorontsov, Cerny, R{\'e}gnier, Di~Jorio, Pal,
  Lapointe, Vandenbroucke-Menu, Turcotte, Kadoury and Tang}]{b24}
\bibinfo{author}{Vorontsov, E.}, \bibinfo{author}{Cerny, M.},
  \bibinfo{author}{R{\'e}gnier, P.}, \bibinfo{author}{Di~Jorio, L.},
  \bibinfo{author}{Pal, C.J.}, \bibinfo{author}{Lapointe, R.},
  \bibinfo{author}{Vandenbroucke-Menu, F.}, \bibinfo{author}{Turcotte, S.},
  \bibinfo{author}{Kadoury, S.}, \bibinfo{author}{Tang, A.},
  \bibinfo{year}{2019}.
\newblock \bibinfo{title}{Deep learning for automated segmentation of liver
  lesions at ct in patients with colorectal cancer liver metastases}.
\newblock \bibinfo{journal}{Radiology. Artificial intelligence}
  \bibinfo{volume}{1}.
%Type = Inproceedings
\bibitem[{Vorontsov et~al.(2018)Vorontsov, Tang, Pal and Kadoury}]{b25}
\bibinfo{author}{Vorontsov, E.}, \bibinfo{author}{Tang, A.},
  \bibinfo{author}{Pal, C.}, \bibinfo{author}{Kadoury, S.},
  \bibinfo{year}{2018}.
\newblock \bibinfo{title}{Liver lesion segmentation informed by joint liver
  segmentation}, in: \bibinfo{booktitle}{2018 IEEE 15th International Symposium
  on Biomedical Imaging (ISBI 2018)}, \bibinfo{organization}{IEEE}. pp.
  \bibinfo{pages}{1332--1335}.
%Type = Article
\bibitem[{Wang et~al.(2016)Wang, Naghavi, Allen, Barber, Bhutta, Carter, Casey,
  Charlson, Chen, Coates et~al.}]{b0}
\bibinfo{author}{Wang, H.}, \bibinfo{author}{Naghavi, M.},
  \bibinfo{author}{Allen, C.}, \bibinfo{author}{Barber, R.M.},
  \bibinfo{author}{Bhutta, Z.A.}, \bibinfo{author}{Carter, A.},
  \bibinfo{author}{Casey, D.C.}, \bibinfo{author}{Charlson, F.J.},
  \bibinfo{author}{Chen, A.Z.}, \bibinfo{author}{Coates, M.M.}, et~al.,
  \bibinfo{year}{2016}.
\newblock \bibinfo{title}{Global, regional, and national life expectancy,
  all-cause mortality, and cause-specific mortality for 249 causes of death,
  1980--2015: a systematic analysis for the global burden of disease study
  2015}.
\newblock \bibinfo{journal}{The lancet} \bibinfo{volume}{388},
  \bibinfo{pages}{1459--1544}.
%Type = Inproceedings
\bibitem[{Wang et~al.(2018)Wang, Girshick, Gupta and He}]{b39}
\bibinfo{author}{Wang, X.}, \bibinfo{author}{Girshick, R.},
  \bibinfo{author}{Gupta, A.}, \bibinfo{author}{He, K.}, \bibinfo{year}{2018}.
\newblock \bibinfo{title}{Non-local neural networks}, in:
  \bibinfo{booktitle}{Proceedings of the IEEE conference on computer vision and
  pattern recognition}, pp. \bibinfo{pages}{7794--7803}.
%Type = Article
\bibitem[{Wang et~al.(2019)Wang, Dou, Hu, Zhu, Yang, Xu, Qin, Heng, Wang and
  Ni}]{b21}
\bibinfo{author}{Wang, Y.}, \bibinfo{author}{Dou, H.}, \bibinfo{author}{Hu,
  X.}, \bibinfo{author}{Zhu, L.}, \bibinfo{author}{Yang, X.},
  \bibinfo{author}{Xu, M.}, \bibinfo{author}{Qin, J.}, \bibinfo{author}{Heng,
  P.A.}, \bibinfo{author}{Wang, T.}, \bibinfo{author}{Ni, D.},
  \bibinfo{year}{2019}.
\newblock \bibinfo{title}{Deep attentive features for prostate segmentation in
  3d transrectal ultrasound}.
\newblock \bibinfo{journal}{IEEE transactions on medical imaging}
  \bibinfo{volume}{38}, \bibinfo{pages}{2768--2778}.
%Type = Inproceedings
\bibitem[{Zhang et~al.(2021a)Zhang, Liu and Hu}]{b45}
\bibinfo{author}{Zhang, Y.}, \bibinfo{author}{Liu, H.}, \bibinfo{author}{Hu,
  Q.}, \bibinfo{year}{2021}a.
\newblock \bibinfo{title}{Transfuse: Fusing transformers and cnns for medical
  image segmentation}, in: \bibinfo{booktitle}{International Conference on
  Medical Image Computing and Computer-Assisted Intervention},
  \bibinfo{organization}{Springer}. pp. \bibinfo{pages}{14--24}.
%Type = Inproceedings
\bibitem[{Zhang et~al.(2021b)Zhang, Peng, Peng, Huang, Tong, Lin, Li, Chen,
  Chen, Hu and Peng}]{b36}
\bibinfo{author}{Zhang, Y.}, \bibinfo{author}{Peng, C.}, \bibinfo{author}{Peng,
  L.}, \bibinfo{author}{Huang, H.}, \bibinfo{author}{Tong, R.},
  \bibinfo{author}{Lin, L.}, \bibinfo{author}{Li, J.}, \bibinfo{author}{Chen,
  Y.W.}, \bibinfo{author}{Chen, Q.}, \bibinfo{author}{Hu, H.},
  \bibinfo{author}{Peng, Z.}, \bibinfo{year}{2021}b.
\newblock \bibinfo{title}{Multi-phase liver tumor segmentation with spatial
  aggregation and uncertain region inpainting}, in:
  \bibinfo{booktitle}{International Conference on Medical Image Computing and
  Computer-Assisted Intervention}, \bibinfo{organization}{Springer}. pp.
  \bibinfo{pages}{68--77}.
%Type = Inproceedings
\bibitem[{Zhao et~al.(2017)Zhao, Shi, Qi, Wang and Jia}]{b30}
\bibinfo{author}{Zhao, H.}, \bibinfo{author}{Shi, J.}, \bibinfo{author}{Qi,
  X.}, \bibinfo{author}{Wang, X.}, \bibinfo{author}{Jia, J.},
  \bibinfo{year}{2017}.
\newblock \bibinfo{title}{Pyramid scene parsing network}, in:
  \bibinfo{booktitle}{Proceedings of the IEEE conference on computer vision and
  pattern recognition}, pp. \bibinfo{pages}{2881--2890}.
%Type = Article
\bibitem[{Zhao et~al.(2021)Zhao, Li, Xiao, Accorsi, Marshall, Cossetto, Kim,
  McCarthy, Dawson, Knezevic et~al.}]{b49}
\bibinfo{author}{Zhao, J.}, \bibinfo{author}{Li, D.}, \bibinfo{author}{Xiao,
  X.}, \bibinfo{author}{Accorsi, F.}, \bibinfo{author}{Marshall, H.},
  \bibinfo{author}{Cossetto, T.}, \bibinfo{author}{Kim, D.},
  \bibinfo{author}{McCarthy, D.}, \bibinfo{author}{Dawson, C.},
  \bibinfo{author}{Knezevic, S.}, et~al., \bibinfo{year}{2021}.
\newblock \bibinfo{title}{United adversarial learning for liver tumor
  segmentation and detection of multi-modality non-contrast mri}.
\newblock \bibinfo{journal}{Medical Image Analysis} \bibinfo{volume}{73},
  \bibinfo{pages}{102154}.

\end{thebibliography}

\end{document}